\documentclass[option]{webofc}
\usepackage[varg]{txfonts}   
\usepackage{amssymb}
\usepackage{amsmath}
\usepackage{hyperref} 
\hypersetup{colorlinks=true, linkcolor=blue, citecolor=blue, urlcolor=blue}

\def\eq#1{(\ref{#1})}
\def\nonu{\nonumber}
\def\bm#1{\text{\boldmath$#1$}}
\def\be{\begin{equation}}
\def\ee{\end{equation}}
\def\bea{\begin{eqnarray}}
\def\eea{\end{eqnarray}}
\def\bean{\begin{eqnarray*}}
\def\eean{\end{eqnarray*}}
\def\gsim{\mathrel{\rlap{\lower0.25em\hbox{$\sim$}}\raise0.2em\hbox{$>$}}} 
\def\lsim{\mathrel{\rlap{\lower0.25em\hbox{$\sim$}}\raise0.2em\hbox{$<$}}}

\newcommand{\Nc}{N_{\rm c}}

\newcommand{\im}{\mathop{\mbox{Im}}}

\newcommand{\CF}{C_{_{\rm F}}}
\newcommand{\Nf}{n_{\!f}}
\newcommand{\muB}{\mu_{\rm B}}
\newcommand{\nF}{n_{\rm F}}
\newcommand{\nB}{n_{\rm B}}
\newcommand{\rhoV}{\rho_{\rm V}}
\newcommand{\rhoH}{\rho_{\rm H}}
\renewcommand{\vec}{\bm}
  \newcommand{\sumint}[1]{{\hbox{$\displaystyle\sum$}\!\!\!\!\!\!\!\!\int\,}_{\!\!\!\!\raise-0.5ex\hbox{$\scriptstyle{#1}$}}}

\begin{document}

\title{Shedding light on thermal photon and dilepton production}

\author{\firstname{Greg} \lastname{Jackson}\inst{1}\fnsep\thanks{\email{
  gsj6@uw.edu
    }} 
}

\institute{
Institute for Nuclear Theory, Box 351550, University of Washington, Seattle, WA 98195-1550, United States}

\abstract{%
Electromagnetic radiation from the quark-gluon plasma (QGP) is an important observable 
to be considered in heavy ion collision experiments. 
I will provide an update on recent advancements 
from perturbation theory and quenched lattice simulations. 
The resummed next-to-leading order (NLO) emission rate has recently been decomposed into 
transverse and longitudinal components, and extended to non-zero baryon chemical potential. 
The associated spectral function has also been tested against the Euclidean 
correlator, 
for continuum-extrapolated lattice data (at $\muB=0$). 
}
\maketitle
\section{Introduction}
\label{intro}

Quarks undergoing acceleration in the deconfined state 
of QCD matter can generate electromagnetic radiation,
with those photons that are off-shell subsequently decaying into lepton-antilepton pairs.
Therefore, both the real photon spectrum and dilepton invariant mass distribution
can provide access to properties of the hot quark-gluon plasma (QGP) 
that exists in heavy ion collision experiments
\cite{McLerran1984,Weldon1990,Gale1990}.
In this report, I will 
show that the spectral function can be constrained by lattice data and
discuss how the presence of a baryon chemical potential, $\muB$,
impacts the production of photons and dileptons.
While the latter has been examined for real photons \cite{Gervais2012},
we present new results away from the light cone.
This involves properly understanding how $\muB$ enters the strict NLO 
computation, the so-called LPM effect (at low invariant masses),
and how to smoothly interpolate between the two regimes as originally
advocated in ref.~\cite{Ghisoiu2014}.

To start, we fix the notation and denote the temperature by $T\,$, 
the quark chemical potential by $\mu\,$ and
the energy and momentum with respect to the plasma rest frame 
of the lepton pair by $\omega$ and $\bm k$ respectively.
In chemical equilibrium,
$\mu = \frac13 \muB$ 
and thermal averages are calculated from 
$\langle ... \rangle = {\rm Tr}\big[\, \hat \varrho\, (...)\,\big]$ 
with the density matrix $\hat \varrho = {\cal Z}^{-1} e^{-(\hat H - \mu \hat Q)/T}$
\cite{basics}.
Emission rates can then be derived from an associated spectral function.
In this case, the relevant spectral function is given by
the imaginary part of the current-current
correlation function, evaluated at the energy $k_0 = \omega + i 0^+$,
namely
\be
 \rho_{\mu\nu}(\omega, \bm k) = 
 \im \big[ \Pi_{\mu\nu}(K)
     \big] 
 \,, \label{cut}
\ee
where $K = (k_0,\vec{k})$
and the correlation function being given by\footnote{%
  An overall minus sign appears in \eq{Pi_munu}
  for sake of convenience. 
}
\be
 \Pi^{ }_{\mu\nu}(K)
 \; \equiv \; 
 -\, 
 \int_0^{1/T} \!\! {\rm d} \tau
 \int_{\vec{x}} \; 
 e^{k_0\tau + i\, \vec{k}\cdot\vec{x} } \,
 \bigg\langle\,
   J_\mu 
   \big(t,\vec{x}\!\,\big) \, 
   J_\nu
   \big(0,\vec{0}\!\,\big) 
 \,\bigg\rangle
 \, , \quad J_\mu \equiv \bar\psi \gamma_\mu \psi
 \;. \label{Pi_munu}
\ee
The thermal average is taken on a volume
with periodic temporal extent $\tau \in (0,T^{-1})$ and 
$k_0$ is a bosonic Matsubara frequency 
$k_0 = i\, 2\pi z T$ with 
$z \in \mathbb{Z}$.

With these definitions, 
the differential photon rate involves the spectral function 
$\rhoV \equiv \rho_\mu^{\ \mu}$ 
for both $i$) real and $ii$) virtual photons.
In case $i$), $\omega = |\vec{k}| \equiv k$
and case $ii$) provides the dilepton rate
 with the invariant mass $M \equiv \sqrt{ K^2}$
 that should be above the threshold to
 form the pair: $K^2 = \omega^2 - k^2 > 4m_\ell^2\,$.
One might also consider $iii$) deep inelastic scattering on a QGP target,
which would involve $\rhoV$ for timelike virtualities \cite{harvey_DIS}.
Although case $iii$) may not be experimentally accessible,
there is another good reason to pursue $K^2<0\,$:
Knowing 
the spectral function at fixed $k$ for {\em all} $\omega$,
enables one to calculate the imaginary time  correlation function
and thus connect with non-perturbative lattice measurements (at $\mu=0$).

\begin{figure}[h]
\centering
\includegraphics[width=9cm,clip]{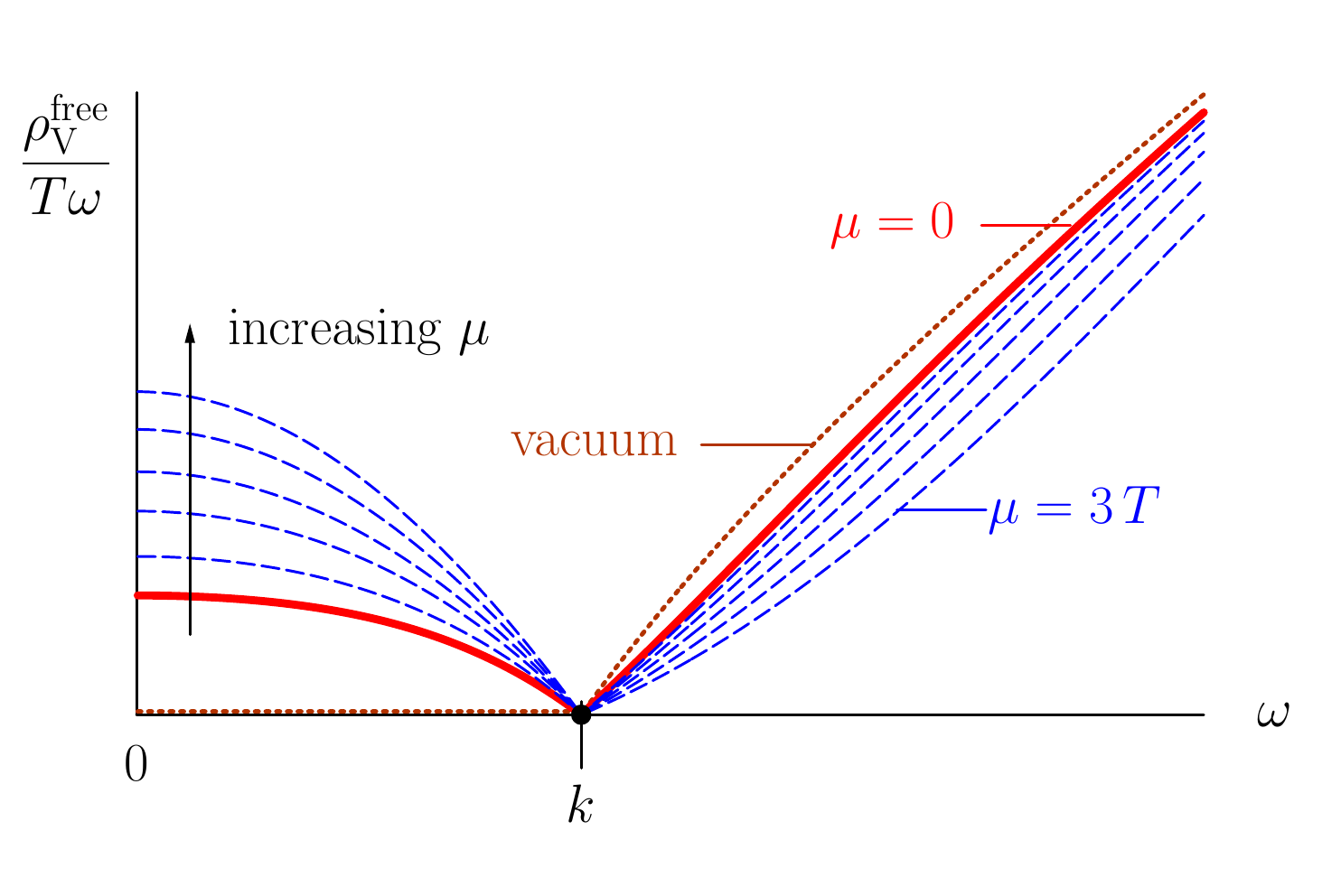}
  \vspace{-.5cm}
\caption{
  Sketch of the 
  free spectral function ($\alpha_s=0$), 
  with the $\mu=0$ result (solid) and
  the impact of $\mu>0$ also shown (dashed).
  The limit $T,\,\mu \to 0$  
  of eq.~\eq{rhoV_LO} is the vacuum result (dotted).
  }
\label{fig-1}
\end{figure}

The gross features of $\rhoV$ can be understood from the
leading-order (LO) process ${q\bar q \to \gamma^\star}$, i.e. $\alpha_s = 0$.
For non-zero $\mu$, the 
result was determined in ref.~\cite{Dumitru1993} for $K^2 > 0\,$.
In general, for any $\omega\,$, the free spectral function is 
given by the strict 1-loop result\footnote{%
  Setting $\mu=0$ gives eq.~(2.4) from ref.~\cite{Jackson2019}.}
\bea
 \left.\rho_{\rm V}\right|^{\rm strict}_{\rm 1-loop} & = &
  \frac{ \Nc K^2 }{4 \pi }
  \bigg\{ \,
 \sum_{\nu = \pm \mu} 
 \frac{T}{k} \ln \bigg[ \frac{1 + e^{\nu-\frac12(\omega+k)/T}}{1+ e^{\nu-\frac12|\omega-k|/T}} \bigg] + \Theta \big( K^2 \big) \,
  \bigg\}
\;, \label{rhoV_LO} 
\eea
where $\Theta$ denotes the Heaviside step function
and $\Nc$ is the number of colours.
This is depicted in fig.~\ref{fig-1}, assuming non-zero $T$.
The vanishing of $\rhoV$ for $\omega = k$ is readily understood
from kinematics
and the $\mu$-dependence stems from the relative 
enhancement and depletion of quarks and antiquarks respectively (for $\mu>0$).

\section{Weak coupling QCD corrections}

Corrections to \eq{rhoV_LO} may be computed in perturbation theory,
however the structure of the expansion in $\alpha_s$ depends
on the (parametric) value of  $K^2$.
Away from the light cone, $|K^2| \gsim (\pi T)^2\,$,  the 2-loop corrections may be calculated 
directly and the NLO terms are ${\cal O}(\alpha_s)$ \cite{Laine2013vpa}.
However, for small $K^2$ as the free
result \eq{rhoV_LO} gets kinematically suppressed (and, in particular vanishes for $K^2=0$) 
implying that the QCD `corrections' actually represent the first non-trivial approximation to the 
real photon rate.
For $K^2 \lsim (gT)^2$ certain diagrams
need to be resummed to obtain a meaningful result, which is motivated on physics grounds
to describe thermal screening \cite{Braaten1990,Kapusta1991,Baier1991}
in addition to the Landau-Pomeranchuk-Migdal (LPM)  effect
\cite{agz,agmz,Arnold2001ba,Arnold2001ms}.
These contributions alter the asymptotic dependence on the strong coupling $\alpha_s$
to a leading-logarithm $\rhoV \sim \alpha_s \ln (1/\alpha_s) T^2$ as $\alpha_s \to 0\,$.

In ref.~\cite{Ghisoiu2014}, a simple procedure to 
interpolate between these two regimes was proposed.
Care is required to avoid double counting when resummation is combined with the strict NLO 
expansion.
A full resummed spectral functions can be defined as 
\bea
 \rhoV |_{\rm NLO}^{\rm resummed}
 \; \equiv \; 
 \rhoV |_{\rm 1-loop}^{\rm strict}
 + 
 \rhoV |_{\rm 2-loop}^{\rm strict}
 + 
 \big( 
 \rhoV |_{\rm LPM}^{\rm full}
 - 
 \rhoV |_{\rm LPM}^{\rm expanded}
 \, \big) 
 \;, \label{resummation} 
\eea
where the subtracted term in parenthesis represent the 1- and 2-loop parts that are included
in the full LPM result (with certain approximations).
For \eq{resummation} to make sense, 
a delicate cancellation must take place around $\omega \simeq k$ so that the result is finite and continuous there~\cite{Jackson2019}.
The details of the LPM `full' and `expanded' will be provided
in sec.~\ref{sec LPM}, where we focus on non-zero baryon density.

In ref.~\cite{Jackson2021}, we studied (with full generality) the types
of interactions that would contribute to strict NLO rates and developed a numerical 
routine for {\em any} combination of particles, masses, chemical potentials and
 a wide class of matrix elements.
For the dilepton rate, it is preferable to use a more tailored approach
which requires a 2-dimensional phase space integration \cite{Jackson2019a}.
The underlying spectral function can be reduced to a set of elementary 
`master integrals' at NLO
(some of which were studied for $\omega >k$ in \cite{Laine2013vpa}),
which are uniformly defined by
\bea
\rho_{abcde}^{(m,n)}
(K) 
&\equiv&
\im\,
\footnotesize\sumint{\, P,Q} \, 
\frac{p_0^m \, q_0^n}{
  [P^2]^a \, [Q^2]^b \, [R^2]^c \, [L^2]^d \, [V^2]^e
}
\, \Bigg|_{R = K-P-Q\, ,\  L = K-P\, ,\  V = K-Q} \ .
\label{I def}
\eea
Functions of this kind provide a basis onto which the
general 2-loop topology (after carrying out the Dirac algebra, etc.)
can be mapped for self energies with external momentum $K$.
In the sum-integrals \eq{I def},\footnote{%
  To be crystal clear, the sum-integrals are (in $4-2\epsilon$ spacetime dimensions)
$$
\sumint{\, P} = \int_{\vec{p}} \, T \sum_{p_0} \;, \qquad
\int_{\vec{p}} = \bigg( \frac{e^\gamma \bar \mu^2}{4\pi} \bigg)^\epsilon 
\int\!\! \frac{d^d p}{(2\pi)^d} \, .
$$
}
$P$ and $Q$ are fermionic momenta with
$p_0 = i (2x+1)\pi T + \mu$ and
$q_0 = i (2y+1)\pi T - \mu$ (where $x,y \in \mathbb{Z}$).
(Recall that $K$ is bosonic, 
thus $R=K-P-Q$ is also bosonic while $L=K-P$ and $V=K-Q$ are fermionic.)

\section{Non-perturbative constraints}

Although real-time rates are difficult to compute from
numerical Monte Carlo simulations,
the $\tau$ dependence of the integrand in eq.~\eq{Pi_munu}
can be obtained from Euclidean lattices.
The imaginary-time 
correlation function $G_{\mu\nu}(\tau,k)$
is related to the spectral function from \eq{cut} via the integral transform\footnote{%
  In practice, all the spectral functions studied here are antisymmetric in $\omega \to - \omega$
  and therefore only the first term on the right hand side of eq.~\eq{G} contributes.
}
\bea
G_{\mu\nu}(\tau,k) \ =\
\int_0^\infty \frac{{\rm d}\omega}{2\pi}
\!\!\! &\Bigg\{& \!\!\!
\Big(\, \rho_{\mu\nu} (\omega,k) - \rho_{\mu\nu} (-\omega,k) \,\Big) 
\frac{ \cosh\big[\omega(\frac1{2T}-\tau)\big] }{\sinh[\frac1{2T} \omega]} 
\nonu \\
\!\!\! &+& \!\!\!
\Big(\, \rho_{\mu\nu} (\omega,k) + \rho_{\mu\nu} (-\omega,k) \,\Big)
\frac{ \sinh\big[\omega(\frac1{2T}-\tau)\big] }{\sinh[\frac1{2T} \omega]} \ \,
\Bigg\} \  .
\label{G}
\eea
It is a formidable task to invert \eq{G} and thus obtain the spectral function 
directly from a finite set of sampling points~\cite{inversion,mem}.

Rather than using \eq{G} to obtain $\rhoV$ (from $G_{\rm V}$),
 another spectral function turns out to be convenient:
\bea
\rho_{\rm H} 
&\equiv& \rhoV + 3 \frac{K^2}{k^2} \rho_{00} \, ,
\label{rhoH}
\eea
which is highly suppressed in the ultraviolet
and exactly vanishes in vacuum.
This makes the corresponding Euclidean correlator more sensitive to the
infrared physics of interest \cite{brandt_rhoH}. 
We point out that $\rhoV$ and $\rhoH$ agree on the light cone,
but differ considerably for $\omega \neq k\,$.
The spectral function \eq{rhoH} satisfies a 
sum rule, $\int_0^\infty {\rm d} \omega \, \omega \, \rho_{\rm H}(\omega,k) = 0\,$,
which supplies additional restrictions on any inversion candidates.
Computing both $\rhoV$ and $\rho_{\rm H}$ 
amounts to determining separately 
the transverse and longitudinal components, thus
specifying the entire tensor $\rho_{\mu\nu}$.

\begin{figure}[t]
\centering
\includegraphics[width=6.2cm,clip]{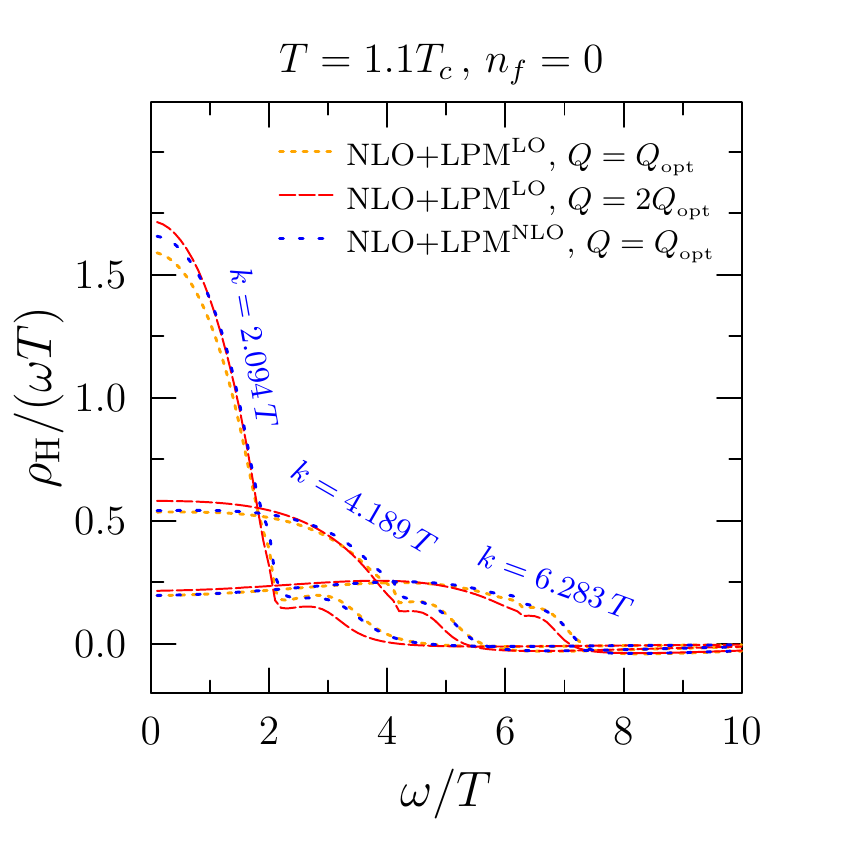}
\!\!\!
\includegraphics[width=6.2cm,clip]{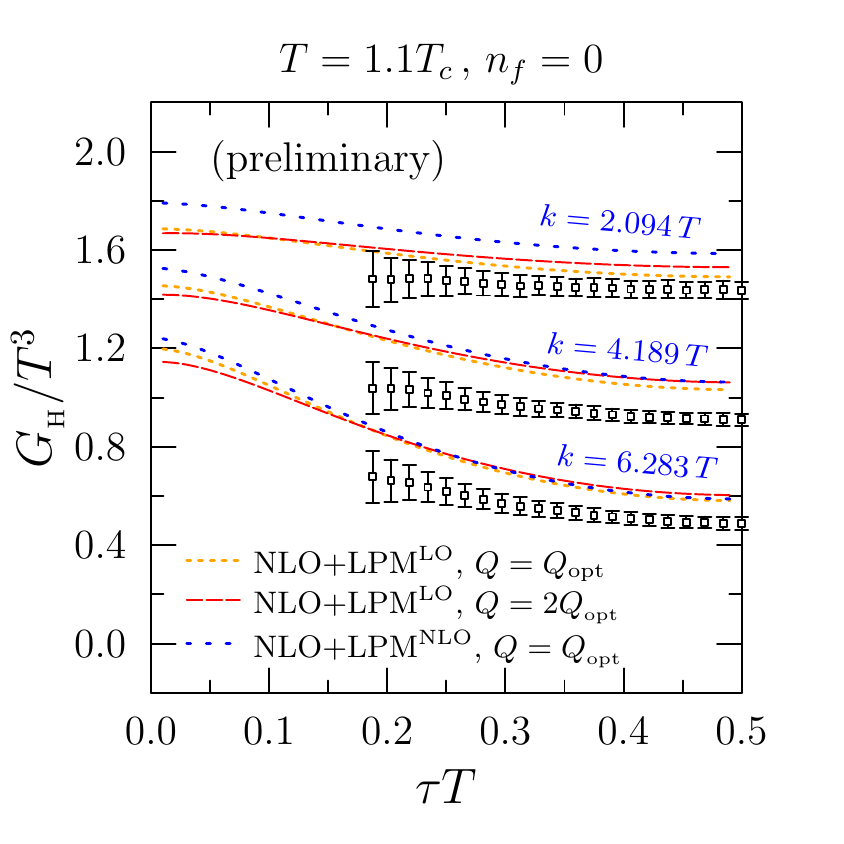}
  \vspace{-.3cm}
\caption{
  Results for $\rhoH$ (left) and the corresponding Euclidean correlation function,
  calculated from eq.~\eq{G}, 
  (right) at $T=1.1T_c$ for $\Nf=0\,$.
  Similar plots for $\rhoV$ (and $G_{\rm V}$) may be found in ref.~\cite{Jackson2019}.
  }
\label{fig-2}
\end{figure}

One may also use \eq{G} to compute $G_{\mu\nu}$ from
models of the spectral function.
In fig.~\ref{fig-2} the perturbative results for $\rhoH$ and $G_{\rm H}$
are shown, compared with 
continuum extrapolated lattice data for quenched QCD from ref.~\cite{constraints}.
The various curves show different choices of the scale in the running coupling, $Q$,
as well as including the NLO part of the LPM computation \cite{lpm_nlo}.
(Further details may be found in refs.~\cite{Jackson2019,phd}.)

\section{Beyond leading-order: strict NLO}
\label{sec NLO}

The strict NLO result for $\rhoV$ can be 
expressed as a linear combination of the master integrals, 
defined by eq.~\eq{I def}.
Evaluating the diagrams, we obtain the result
\bea
 \left.\rhoV\right|_{\rm NLO}^{(g^2)} & = &
\label{masters, mn}\\
8(1-\epsilon) g^2 \CF \Nc  
\!\!\! & \bigg\{ &
(1-\epsilon)K^2
\Big( 
  \rho_{11020}^{(0,0)} + \rho_{11002}^{(0,0)} 
- \rho_{10120}^{(0,0)} - \rho_{01102}^{(0,0)} \Big)
+ \rho_{11010}^{(0,0)}  
+ \rho_{11001}^{(0,0)}  
\nonu \\
&+& 2\epsilon \, \rho_{11100}^{(0,0)} 
\  +\ 2 \frac{K^2}{k^2} \rho_{11011}^{(1,1)} 
\ -\ \tfrac12 K^2 
\Big(\, \frac{\omega^2}{k^2} + 3+2\epsilon \,\Big) 
\rho_{11011}^{(0,0)}
\nonu \\
&-& 
 (1-\epsilon) \Big( \rho_{1111(-1)}^{(0,0)} + \rho_{111(-1)1}^{(0,0)}  \Big)
+ 2K^2  \Big( \rho_{11110}^{(0,0)} + \rho_{11101}^{(0,0)} \Big)
- K^4 \rho_{11111}^{(0,0)} 
\ \bigg\}  \ ,\nonu 
\eea
where $\CF = (\Nc^2-1)/(2\Nc)\,$.
Above, the limit $\epsilon \to 0$ is implied because
some of the master integrals have $1/\epsilon$-contributions stemming from
their vacuum parts.
Note that the spectral function is symmetric in the simultaneous
exchanges:
$a \leftrightarrow b\,$, $d \leftrightarrow e$ and $m \leftrightarrow n\,$ 
for the master integrals.
Consequently, the result will be unchanged by $\mu \to - \mu\,$.
In the case where $\mu = 0\,$, the additional symmetry 
$\rho_{abcde}^{(m,n)}
\to \rho_{baced}^{(n,m)}$ leads to the same 
decomposition as in ref.~\cite{Jackson2019,Jackson2019a}.

An important cross-check of the result \eq{masters, mn} (besides the obvious, gauge invariance etc.)
can be found 
within the {\em hard thermal loop} (HTL) approximation, for which 
the master integrals can be computed in closed form. 
The HTL limit is given by the small-$K$ behaviour of 
$\Pi_{\mu\nu}$ and the 1-loop result is well known \cite{basics}.
Recently, the 2-loop HTL photon self energy was 
computed for a hot and dense QED plasma in ref.~\cite{Gorda2022}.
We restate the outcome here,
in a way that is compatible with eq.~\eq{Pi_munu}, 
\bea
\Pi_{\rm V}^{\rm HTL} 
\!\!\! &=& \!\!
- \, \bigg( \tfrac13 T^2 + \frac{\mu^2}{\pi^2} \bigg)
+ \frac{e^2}{8\pi^2} \bigg( T^2 + \frac{\mu^2}{\pi^2} \bigg) 
  \bigg( 1 + \frac{\omega}{k} 
  L 
  \bigg) 
+ \frac{e^2}{4\pi^2} \, \frac{\mu^2}{\pi^2} \bigg( 1 - \frac{\omega^2}{k^2} \bigg)
  \bigg( 1 - \frac{\omega}{2k} 
  L 
  \bigg)^2 \ ,
  \nonu \\
  \label{htl}
\eea
where $L = \ln \frac{\omega + k + i0^+}{\omega - k + i0^+}\,$.
This result can be transcribed to the present 
case by replacing ${e^2 \to g^2 \CF \Nc}$
in eq.~\eq{htl},
so that the resulting spectral function should coincide with  
the strict NLO version of $\rhoV$ \eq{masters, mn} assuming $\omega$ and $k$ are small.
The agreement between the two approaches 
has been verified both analytically and numerically.
Worth mentioning explicitly, is the HTL limit for the 
$\rho_{11011}^{(1,1)}$ master integral.\footnote{%
If $\mu=0\,$, one can prove that 
 $\rho_{11011}^{(1,1)} =  \frac14 \omega^2 \rho_{11011}^{(0,0)}\,$
 which vanishes in the HTL approximation when $\omega$ is soft.
  }
One may readily check that
\bea
\sumint{P,Q} 
\frac{p_0 q_0}{P^2Q^2(K-P)^2(K-Q)^2}
& \approx &
 -\, 
\frac{\mu^2}{4(2\pi)^4} 
\bigg(\, 1 -  \frac{\omega}{2k} L 
\bigg)^2 
\, . 
\eea
This term appears when the strict 2-loop self energy, 
$\Pi_{\rm V}|_{\rm NLO}^{(g^2)}$, is evaluated 
and is entirely responsible for the last term in \eq{htl},
which  contains a new structure involving a squared logarithm (only
present at finite density).

\section{Beyond leading-order: LPM regime}
\label{sec LPM}

The master integrals  
$\rho_{1111(-1)}^{(0,0)}$ and $\rho_{111(-1)1}^{(0,0)}$
from eq.~\eq{masters, mn} each contain a log-divergence as $K^2\to 0^\pm$ \cite{phd}.
This is a signal that resummation is required, and the LPM framework
serves that purpose.
Two important scales enter in the problem: The
Debye mass $m_D$ and the asymptotic quark mass $m_\infty$,
both of which are modified by the chemical potential, {\em viz.}
\bea
m_D^2  \ \equiv \ 
g^2 \bigg[
\Big(\tfrac12 \Nf + \Nc \Big)
\frac{T^2}3 
+
\Nf \, \frac{\mu^2}{2\pi^2} \bigg] \;,
& &
m_\infty^2 \ \equiv \ 
g^2 \, \frac{\CF}4 \bigg( T^2 + \frac{\mu^2}{\pi^2} \bigg) \; ,
\eea
where $\Nf$ is the number of light quark flavours.
Following ref.~\cite{agmz} in impact parameter space, 
the result can be expressed as\footnote{%
  A formulation with better asymptotics (for large $M$) was
  proposed in ref.~\cite{lpm_born_interp}, although we do not use that here.
  }
\bea
 \left. \rhoV \right|^{\rm full}_{\rm LPM} 
 & = &  
 - \frac{\Nc}{\pi}
 \int_{-\infty}^{\infty} \! {\rm d}p \, 
 \bigl[ 1-\nF^{ }(p-\mu)-\nF^{ }(\omega-p+\mu) \bigr]
 \\ \nonu
 & \times &  
 \lim_{{\bm b}\to {\bm 0}} \,  \mathbb{P} \,
 \biggl\{ 
   \frac{K^2
   }{\omega^2}
   \im [g({\bm b})]
  + 
  \frac12 \bigg[ \frac1{p^2} + \frac1{(\omega-p)^2} \bigg]
   \im [\nabla^{ }_{\perp}\cdot {\bm f}({\bm b})] 
 \biggr\} 
 \;, \label{final_from_LPM} \hspace*{5mm}
\eea
where 
$\nF$ is the Fermi-Dirac distribution,
$\mathbb{P}$ stands for the Cauchy principal value and 
$g$ and ${\bm f}$ are Green's functions satisfying
\be
 \bigl( \hat{H} + i 0^+_{ }\bigr) g({\bm b}) \; = \; \delta^{(2)}({\bm b})
 \;, \quad
 \bigl( \hat{H} + i 0^+_{ }\bigr) {\bm f}({\bm b}) 
 \; = \; -\nabla^{ }_{\perp} \delta^{(2)}({\bm b}) 
 \;. 
\ee
The operator $\hat{H}$ acts in the transverse plane,
\be
 \hat{H} = 
  \frac{\omega^{ }
  ( M_{\rm eff}^2 - \nabla_\perp^2 )
  }{2p(\omega^{ }- p)}
 + i g^2 \CF^{ } T
   \int\! \frac{{\rm d}^2 \vec{q}}{(2\pi)^2}
  \bigl( 1 - e^{i {\bm q}\cdot{\bm b}}\bigr)
  \biggl( 
   \frac{1}{q^2} - \frac{1}{q^2 + m_D^2}
  \biggr)
 \;, \label{hatH}
\ee
where $M_{\rm eff}^2 \equiv m_\infty^2 - \frac{p(\omega-p)}{\omega^2}M^2\,$.

\begin{figure}
\centering
\sidecaption
\includegraphics[width=7.5cm,clip]{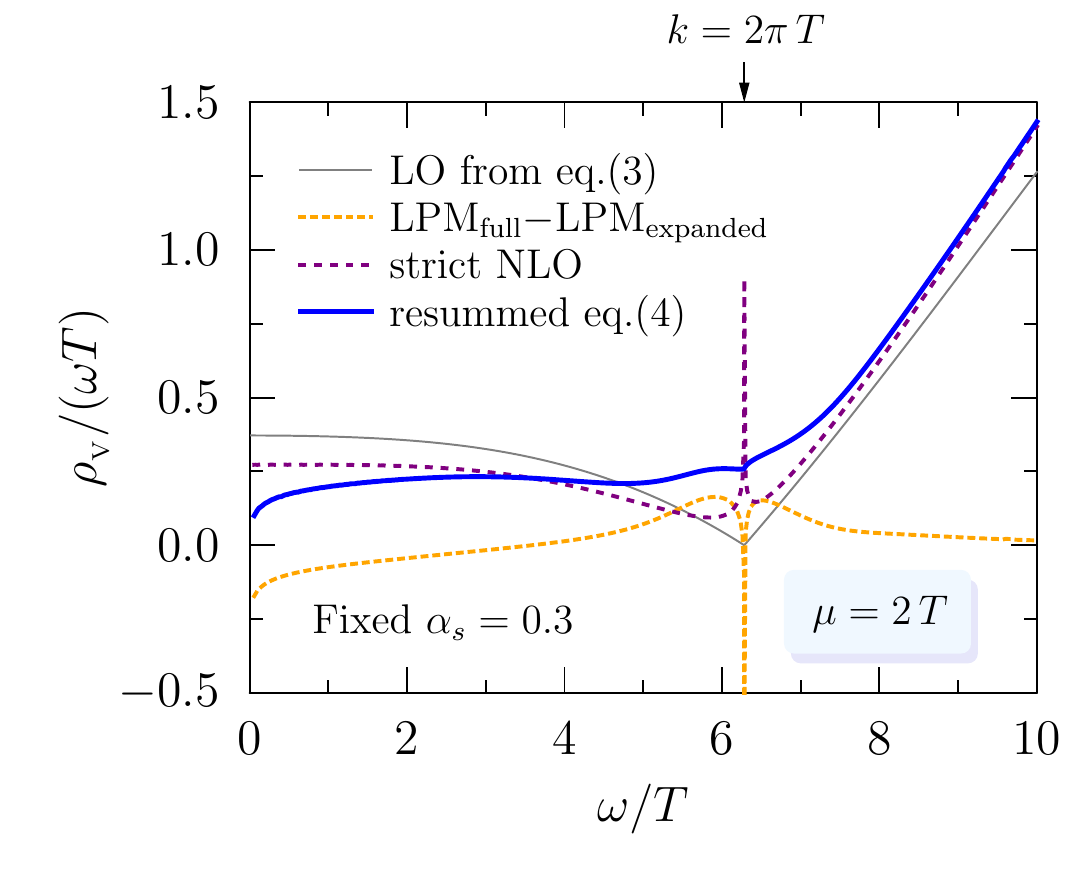}
\quad
\caption{
  The vector channel spectral function, plotted as a function of $\omega$ for $k=2\pi T$
  and with $\alpha_s = 0.3\,$. 
  This illustrates both the strict loop expansion from sec.~\ref{sec NLO} (dashed) 
  and the relevant LPM part from sec.~\ref{sec LPM} (dotted)
  where the log-divergence for $\omega \approx k$ from \eq{log div} is evident.
  Crucially, the prescription of eq.~\eq{resummation} gives a result (solid) that
  is both finite and continuous at the light cone.
  (For comparison, the free result is also shown.)
  }
\label{fig-3}
\end{figure}

In order to combine the LPM and NLO results, 
we also need to naively expand the LPM results up to ${\cal O}(g^2)$
and remove double counting {\em \`{a} la} eq.~\eq{resummation}.
At zeroth order in $g$,  
the expression becomes
\bea
 \rhoV \big|_{\rm LPM}^{(g^0)}
 \; = \; 
  \frac{ \Nc M^2 }{4 \pi }
  \bigg\{ \,
 \sum_{\nu = \pm \mu} 
 \frac{T}{\omega} \ln \bigg[ \frac{1 + e^{(\nu-\omega)/T}}{1+ e^{\nu/T}} \bigg] + \Theta \big( K^2 \big) \,
  \bigg\}
\;, \label{LPM_LO} 
\eea
which matches \eq{rhoV_LO} for $\omega \simeq k\,$.
The corrections of ${\cal O}(g^2)$ 
are proportional to $m_\infty^2\,$.
As in the $\mu=0$ case \cite{Jackson2019}, 
the  spectral function $\rhoV$ contains a
log-divergence plus a finite part:
\bea
 \rhoV \big|^{(g^2)}_{\rm LPM} 
 & = & 
 \frac{\Nc^{ } m_\infty^2 }{4\pi}
 \Biggl\{ 
   \biggl[ 1 - \nF(\omega-\mu) - \nF(\omega+\mu) \biggr]
   \biggl( \ln\biggl| \frac{m_\infty^2}{M^2}  \biggr| - 1 \biggr)
   \label{LPM T}
   + {\cal F}(\omega) \, \Biggr\} 
   \label{log div}
\eea
where
\bea
{\cal F}(\omega) &\equiv&
 \biggl[
 \Theta(K^2) \!
 \int_0^{\omega}\! {\rm d}p
 - 
 \Theta(-K^2) 
 \biggl( \, \int_{-\infty}^{0} \! + \int_{\omega}^{\infty} \,\biggr)
 \, {\rm d}p  
 \biggl] 
 \bigg\{
   2\,\frac{1 - \nF(p-\mu) - \nF(\omega-p+\mu)}{\omega}
 \, \nonu \\
 &-& 
   \frac{\nF(-\mu) + 
   \nF(\omega+\mu) -  
   \nF(p-\mu)  -
   \nF(\omega-p+\mu)}{p}
 \nonu \\
 & - &
 \frac{
   \nF(\omega-\mu) +
   \nF(\mu) -
   \nF(p-\mu) -  
   \nF(\omega-p+\mu)   
 }{\omega-p}
 \ \bigg\} \ .
\eea
The log-divergence in \eq{log div} exactly matches that from $\rhoV|_{\rm NLO}^{(g^2)}\,$, 
and the full resummed expression is finite and continuous across the light cone. 
This is illustrated in fig.~\ref{fig-3} at $\mu=2T$
for fixed coupling $\alpha_s=0.3\,$.
(We have also verified this cancellation analytically.)
Although not visible from fig.~\ref{fig-3}, the presence of $\mu$
enhances the LPM rate due to a larger $m_\infty$ which sets 
the overall scale. 
This enhancement counteracts the suppressing effect of $\mu$ in the
1-loop spectral function \eq{rhoV_LO}.

\section{Outlook}

The emission rate of thermal photons and dileptons can be derived from the
same underlying spectral function $\rhoV$, which encodes all orders in $\alpha_s$.
After a long history of computing the perturbative corrections 
in various limits,
there is now sufficient information to interpolate between these 
regimes as suggested by ref.~\cite{Ghisoiu2014}.
The utility of having a model of the spectral function for all $\omega$ 
is that it allows for comparison with lattice data at non-zero momentum.
One may also use the pertubative result to create `mock data' for testing
methods of reconstructing $\rhoV$ from \eq{G}, e.g. the Backus-Gilbert method.

\begin{figure}[h]
\centering
\sidecaption
\includegraphics[width=7.5cm,clip]{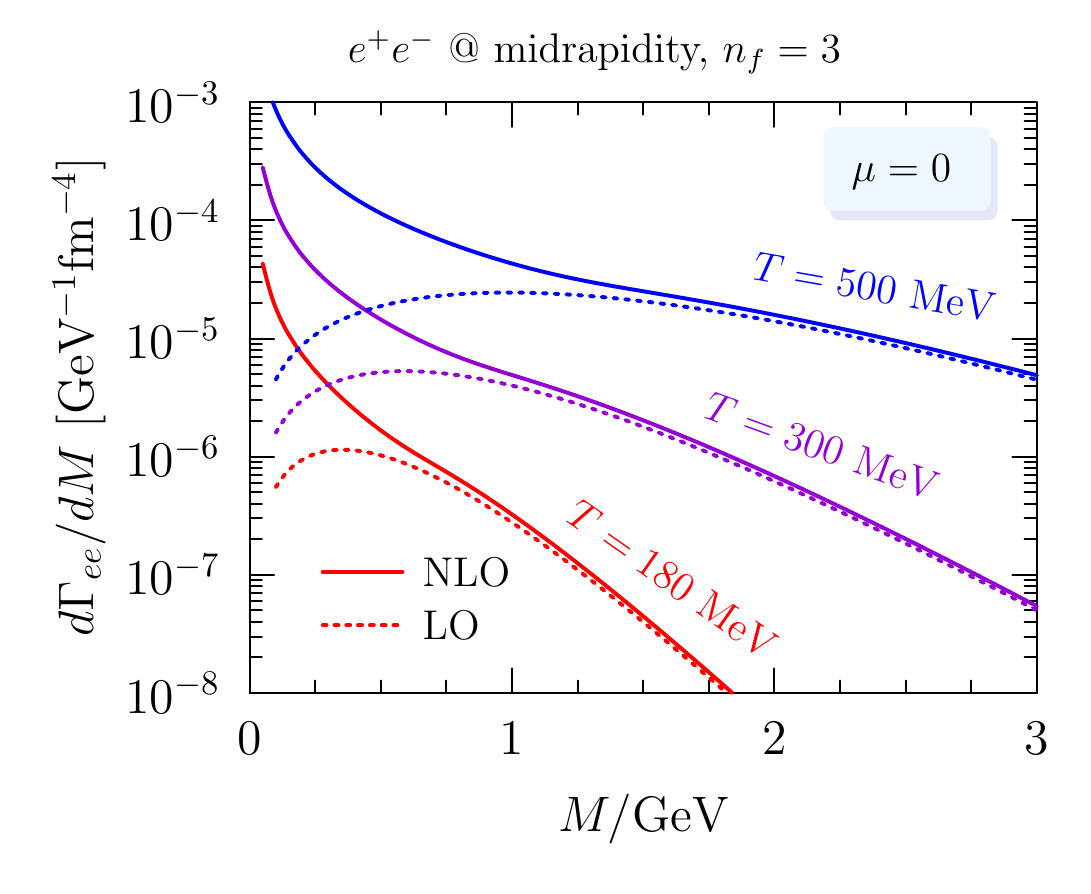}
\quad
\caption{
  Invariant mass distribution of the differential rate to product an
  $e^+e^-$ pair, ${\rm d}\Gamma_{ee} \equiv {\rm d}N/{\rm d}^4X\,$,
  computed from eq.~\eq{dilepton} after converting to hyperbolic coordinates and integrating over the 
  azimuthal angle and $k_\perp$ at midrapidity.
  We show the resummed NLO result from \eq{resummation} (solid) and the LO result from \eq{rhoV_LO} (dotted),
  at the temperatures $T = \{ 180,300,500 \}$~MeV.
  The running of $\alpha_s$ has been implemented as described in ref.~\cite{Jackson2019}.
  }
\label{fig-4}
\end{figure}

A natural next step is to implement the thermal rates calculated from \eq{resummation}
in hydrodynamic simulations of relativistic heavy ion collisions~\cite{jc}.
(Early studies in this direction can be found in ref.~\cite{Burnier2015}.)
For example, the fully differential dilepton rate, with
$\alpha_{\rm em} = e^2/(4\pi)$ and for $\Nf=3\,$ reads
\bea
 \frac{{\rm d} \Gamma_{\ell\bar\ell}\,(\omega,k)}
   {{\rm d}\omega\, {\rm d}^3 \vec{k}} 
  &=&
 2\frac{\alpha_{\rm em}^2  \nB^{ } (\omega)  } 
  {9 \pi^3 M^2} \;
 B \bigg( \frac{m_\ell^2}{M^2} \bigg) \;
 \rhoV^{ }\big(\omega,{k}\big)
 \; ,\label{dilepton}
\eea
where $\nB$ is the Bose distribution function and
 the phase space factor is $B(x) \equiv (1 + 2x) \Theta( 1 -4x ) \sqrt{ 1 -4x }\,$.
The $M$-distribution that follows is shown in fig.~\ref{fig-4} for several temperatures 
(at zero net baryon density) which are expected to be probed in central collisions
at the LHC and RHIC facilities.
Since ${\rm d}\Gamma_{\ell \bar \ell}$ 
represents the rate per unit volume, the result shown still needs to 
be convoluted with the spacetime evolution of the fireball.
This task is left for future work

\section*{Acknowledgements}

Let me express my gratitude to
D.~Bala, 
J.~Churchill, 
C.~Gale, 
J.~Ghiglieri, 
S.~Jeon, 
O.~Kaczmarek and 
M.~Laine
for many helpful discussions and their ongoing collaboration 
on several aspects of this topic.
Furthermore, I thank J.~Ghiglieri for providing the 
LPM$^{\rm NLO}$ data from ref.~\cite{lpm_nlo},
and D.~Bala and O.~Kaczmarek for providing 
the quenched lattice data shown in fig.~\ref{fig-2}.
I am also grateful to T.~Gorda, K.~Sepp\"{a}nen and R.~Paatelainen 
for their assistance in cross-checking these results in 
the HTL limit \cite{Gorda2022}.
This work was supported by the U.S. Department of Energy (DOE) under grant No.~DE-FG02-00ER41132. 

\bibliography{references}

\end{document}